\documentclass[[journal,12pt,onecolumn,draftclsnofoot,]{IEEEtran}
\IEEEoverridecommandlockouts
\usepackage{url}
\usepackage{microtype}
\usepackage{cite}
\usepackage{tumcolor}
\usepackage{amsmath,amssymb,amsfonts}
\usepackage{algorithmic}
\usepackage{graphicx}
\usepackage{textcomp}
\usepackage{bm}
\usepackage{tikz}
\usepackage{subcaption}
\usepackage{pgfplots}
\usepackage[bookmarks=false]{hyperref}
\hypersetup{
	colorlinks=true,       % false: boxed links; true: colored links
	linkcolor=black,          % color of internal links (change box color with linkbordercolor)
	citecolor=black,        % color of links to bibliography
	filecolor=black,         % color of file links
	urlcolor=black        % color of external links
}
\def\BibTeX{{\rm B\kern-.05em{\sc i\kern-.025em b}\kern-.08em
    T\kern-.1667em\lower.7ex\hbox{E}\kern-.125emX}}

\newcommand{\op}[1]{{\operatorname{#1}}}
\newcommand{\uproman}[1]{\uppercase\expandafter{\romannumeral#1}}
\newcommand{\tr}{\operatorname{tr}}
\newcommand{\vect}{\operatorname{vec}}
\newcommand{\diag}{\operatorname{diag}}

\newcommand{\del}{_{\bm{\delta}}}
\newcommand{\deli}{_{\bm{\delta}_i}}
\newcommand{\eye}{\bm{\op{I}}}
\newcommand{\B}[1]{\bm{#1}}
\newcommand{\inv}{^{-1}}

\newcommand{\h}{^{\op H}}
\newcommand{\T}{^{\op T}}
\newcommand{\siginv}{\frac{1}{\sigma^2}}

\newcommand{\Chat}{\widehat{\B C}}
\newcommand{\chat}{\hat{\B c}}

\newcommand{\GE}{_{\text{GE}}}

\newcommand{\bdelt}{\B\delta}
\newcommand{\Wdel}{\B W\del}
\newcommand{\Cdel}{\B C\del}

\newcommand{\Wha}{\widehat{\B W}}
\newcommand{\dB}{$\operatorname{dB}$}
\newcommand{\diagblock}{\operatorname{diablk}}
\newcommand{\cbar}{\bar{\B c}}

\usepackage[acronym,shortcuts]{glossaries}
\newacronym{blmmse}{BLMMSE}{Bussgang LMMSE}
\newacronym{cnn}{CNN}{convolutional neural network}
\newacronym{dft}{DFT}{discrete Fourier transform}
\newacronym{em}{EM}{expectation-maximization}
\newacronym{emiht}{EM-IHT}{EM algorithm with IHT}
\newacronym{iht}{IHT}{iterative hard thresholding}
\newacronym{lmmse}{LMMSE}{linear minimum mean square error}
\newacronym{mmse}{MMSE}{minimum mean squared error}
\newacronym{relu}{ReLU}{rectified linear unit}
\newacronym{snr}{SNR}{signal-to-noise ratio}
\newacronym{adc}{ADC}{analog-to-digital converter}
\newacronym{BS}{BS}{base station}
\newacronym{MS}{MS}{mobile station}
\newacronym{ula}{ULA}{uniform linear array}
\newacronym{ge}{GE}{gridded estimator}
\newacronym{fe}{FE}{fast estimator}
\newacronym{omp}{OMP}{orthogonal matching pursuit}
\newacronym{ml}{ML}{maximum likelihood}
\newacronym{ls}{LS}{least squares}
\newacronym{fft}{FFT}{fast Fourier transform}
\newacronym{flop}{FLOP}{floting point operation}
\newacronym{ce}{CE}{channel estimation}

\newtheorem{assumption}{Assumption}

\begin{document}

\title{A Low-Complexity MIMO Channel Estimator with Implicit Structure of a Convolutional Neural Network\\
	\thanks{The authors are with the Professur f\"ur Methoden der Signalverarbeitung, Technische Universit\"at M\"unchen, Munich, 80333, Germany (e-mail:  \{benedikt.fesl, michael.koller, nurettin.turan, utschick\}@tum.de).}
\thanks{This work was funded by Huawei Sweden Technologies AB, Lund.}
\thanks{
	\vspace{0.2cm}
	
\noindent\fbox{%
	\parbox{\textwidth}{%
		\copyright This work has been submitted to the IEEE for possible publication. Copyright may be transferred without notice, after which this version may no longer be accessible.
	}%
}
}
}

\author{
	\IEEEauthorblockN{Benedikt Fesl, Nurettin Turan, Michael Koller, and Wolfgang Utschick, \IEEEmembership{Fellow, IEEE}}
%	\IEEEauthorblockA{Professur f\"ur Methoden der Signalverarbeitung, Technische Universit\"at M\"unchen, 80290 Munich, Germany\\
%		Email: \{benedikt.fesl, michael.koller, nurettin.turan, utschick\}@tum.de}
%	%	Email: \small{\texttt{\{benedikt.fesl, michael.koller, nurettin.turan, utschick\}@tum.de}}}
}

\maketitle

\begin{abstract}
	A low-complexity convolutional neural network estimator which learns the minimum mean squared error channel estimator for single-antenna users was recently proposed. We generalize the architecture to the estimation of MIMO channels with multiple-antenna users and incorporate complexity-reducing assumptions based on the channel model. Learning is used in this context to combat the mismatch between the assumptions and real scenarios where the assumptions may not hold. We derive a high-level description of the estimator for arbitrary choices of the pilot sequence. It turns out that the proposed estimator has the implicit structure of a two-layered convolutional neural network, where the derived quantities can be relaxed to learnable parameters. 
	We show that by using discrete Fourier transform based pilots the number of learnable network parameters decreases significantly and the online run time of the estimator is reduced considerably, where we can achieve linearithmic order of complexity in the number of antennas. 
	Numerical results demonstrate performance gains compared to state-of-the-art algorithms from the field of compressive sensing or covariance estimation of the same or even higher computational complexity. The simulation code is available online.
\end{abstract}

\begin{IEEEkeywords}
	Channel estimation, massive MIMO, machine learning, neural networks, spatial channel model.
\end{IEEEkeywords}

\IEEEpeerreviewmaketitle

\section{Introduction}
Accurate \ac{ce} is a key aspect in modern communication technologies such as mm-wave communications \cite{6951995} and cellular massive MIMO systems \cite{5595728}. As with increasing system complexity the estimation becomes more difficult, new approaches are necessary which combine reasonable performance together with low complexity. Recently, machine learning based approaches gained a lot of interest which use simulated or measured data to obtain channel estimators \cite{8400482} - \nocite{8353153}\nocite{Yang2019}\nocite{Ha2021}\cite{Marinberg2020}.
%Nowadays, many neural network based \ac{ce} approaches exist (cf. \cite{8400482} - \nocite{8353153}\nocite{Yang2019}\nocite{Ha2021}\cite{Marinberg2020}). 
To the best of our knowledge, none of these approaches uses side knowledge from the channel model to derive a suitable low-complexity neural network architecture for MIMO \ac{ce}.

%Nowadays, many neural network based \ac{ce} approaches exist (cf. citations x -y), but to the best of our knowledge there exists no low-complexity channel estimator for MIMO channels that is based on direct incorporation of side knowledge from the model that determines the network architecture.

In \cite{8272484}, a \ac{cnn} channel estimator is derived which is based on the \ac{mmse} estimator and includes assumptions which stem from a spatial channel model, e.g., the 3GPP model \cite{3GPP}.
%The authors in \cite{8272484} derived a \ac{cnn} channel estimator which is based on the \ac{mmse} estimator and includes assumptions which stem from a spatial channel model, e.g., the 3GPP model \cite{3GPP}. 
The \ac{cnn} is further trained on simulated data to mitigate the mismatch of the assumptions in real scenarios \cite{8272484}. This approach is applicable for various antenna array configurations \cite{8815482}, for measurement data \cite{8727213} and also for quantized systems \cite{69420}.
However, the \ac{cnn} estimator is limited to SIMO~systems. 

%In this work, we generalize the \ac{cnn} architecture to MIMO scenarios where, in addition to a multi-antenna \ac{BS}, the user is equipped with several antennas and an arbitrary number of pilots is transmitted. This setup complicates the scenario drastically, as antenna correlations play an important role \cite{1203167} and the choice of the pilots impacts the architecture. 
%We derive the \ac{mmse} channel estimator for this generalized setup. Similar to \cite{8272484}, complexity-reducing assumptions from the spatial channel model can be incorporated which results in a different architecture. 
%%After this, we further investigate the choice of the pilot sequence. 
%With \ac{dft}-based pilots, which are very common in communication systems \cite{1597555}, \cite{1045284}, the number of learnable network parameters can be reduced considerably. In the end we obtain a \ac{cnn} estimator with a complexity of only $\mathcal{O}(SU\log(SU))$ \acp{flop}, where $S$ and $U$ are the numbers of antennas at the \ac{BS} and \ac{MS}, respectively. 
%
%The proposed estimator reduces to the SIMO estimator from \cite{8272484} for a single-antenna user and a single pilot. Note that the resulting \ac{cnn} estimator is valid for a whole class of channel covariance matrices, depending on prior channel parameters.

The contributions of this work are summarized as follows. We first generalize the SIMO estimator from \cite{8272484} to the MIMO case. This requires us to also incorporate the transmission of arbitrary pilot signals.
We derive a conditional MMSE estimator that is based on the assumption that prior parameters exist that determine the channel statistics, but which is not computable without knowlege of this prior. Note that this clearly distinguishes the approach from the standard MMSE estimator.
We show that the MIMO estimator can be decomposed into a what we call diagonal-block sparse structure which is different from the SIMO case. We identify the structure of a \ac{cnn} with 2D circular convolution layers instead of the 1D convolutions in \cite{8272484} by untying \cite{Hershey} the derived quantities in the sparse structure. We justify this extension by investigating insights from the spatial channel model.
By choosing orthogonal, e.g., \ac{dft}-based, pilot sequences we can show a structural simplification of the estimator, which reduces to a diagonal structure and decreases the number of learnable network parameters considerably. The resulting \ac{cnn} estimator is valid for a whole class of channel covariance matrices.

We provide simulation results which verify the reasonability of the approach and that the learning procedure is indeed able to compensate the mismatch of the assumptions in the derivations. The proposed method is compared to 
a method based on \ac{ml} estimation of the covariance matrix and also to the compressive sensing based approach \ac{omp}.
We provide evidence of the superiority of our approach to the existing baseline algorithms due the possibility of training a model-based architecture with low online complexity of linearithmic order. The simulation code can be found in \cite{sim_code_new}.
%We come to the conclusion that our learning-based approach is able to outperform existing methods in estimating MIMO channels with the advantage of having low online complexity of logarithmic order.

%that are present in the model and are dependent on prior channel~parameters.

%We compare the  \ac{cnn} estimator to state-of-the-art approaches, including the compressive sensing algorithm \ac{omp} \cite{7499112}, \cite{681706} and a \ac{ml} method which exploits an approximate structure of the channel covariance matrix \cite{7051818}, \cite{1456695}. The simulation code can be found in \cite{sim_code}.
%Simulation results show that the \ac{cnn} estimator outperforms the reference algorithms, especially in setups which are close to real~scenarios. 

\textbf{Notation:} The transpose and conjugate-transpose of a vector $\B x$ are denoted by $\B x\T$ and $\B x\h$.
The $n\times n$ identity matrix and the $n\times 1$ all-ones vector are denoted by $\eye_n$ and $\B 1_n$. The column-wise vectorization and the trace of a matrix $\B X$ is denoted by $\vect(\B X)$ and $\tr(\B X)$, the Kronecker product by $ \otimes$. A diagonal matrix with diagonal $\B x$ is given by $\diag(\B x)$. We denote two matrices $\B X, \B Y\in\mathbb{C}^{M\times M}$ which are asymptotically equivalent, i.e., $\lim_{M\rightarrow\infty} ||\B X-\B Y ||_F^2 / M = 0$ holds, by $\B X \asymp \B Y$.

\section{Channel and System Model}\label{sec:model}
We consider an uplink scenario, where $N$ pilot signals are transmitted. The \ac{BS} and the \ac{MS} consist of \acp{ula} with $S$ and $U$ antennas, respectively.
%We consider an uplink scenario where the \ac{BS} and \ac{MS} consist of a \ac{ula} with $S$ and $U$ antennas, respectively, where $N$ pilot signals are transmitted. 
The channel is assumed frequency-flat with block-fading such that we get independent observations in each coherence interval.
We investigate a single-snapshot scenario, i.e., the coherence interval of the covariance matrix and of the channel is identical. The signal received at the \ac{BS} is $\B Y = \B H \B X^\prime + \B  Z$
%\begin{equation}\label{eq:system}
%	\mathbb{C}^{S\times N}\ni\B Y = \B H \B X^\prime + \B  Z
%\end{equation}
with the channel matrix $\B H\in\mathbb{C}^{S\times U}$ and the pilot matrix $\B X^\prime\in\mathbb{C}^{U\times N}$.
After vectorization we get
\begin{equation}
	\B y = \B X\B h + \B z \in 	\mathbb{C}^{SN},
\end{equation}
with $\B X = \B X^{\prime,\op T} \otimes \eye_S$ and $\B h = \vect(\B H)\hspace{-0.05cm}\in\hspace{-0.05cm}\mathbb{C}^{SU}$. The noise vector is assumed to be distributed as $\B z \hspace{-0.05cm}\sim \hspace{-0.05cm}\mathcal{N}_{\mathbb{C}}(\B 0,\sigma^2\eye_{SN})$.

We work with the 3GPP spatial channel model \cite{3GPP}, where for a given variable $\bdelt$, the channels are distributed as $\B h |\bdelt\sim \mathcal{N}_{\mathbb{C}}(\B 0,\Cdel)$. 
The vector $\bdelt$ contains angles of arrivals and path gains and follows an unknown distribution $\bdelt\sim p(\bdelt)$.
The covariance matrix $\Cdel$
is assumed to be determined by the transmit- and receive-side spatial correlation matrices, i.e., $\Cdel = \B C_{\bdelt,T} \otimes \B C_{\bdelt,R}$, 
%\begin{equation}\label{eq:kron_cov}
%	\Cdel = \B C_{\bdelt,T} \otimes \B C_{\bdelt,R},
%\end{equation}
which is valid in spatial correlation scenarios \cite{837052}, \cite{1021913}. Each covariance matrix equals to
\begin{equation*}
	\B C_{\bdelt,\{T,R\}} = \int_{-\pi}^{\pi} g(\theta_{\{T,R\}};\bdelt ) \B a(\theta_{\{T,R\}})\B a(\theta_{\{T,R\}})\h \op d\theta_{\{T,R\}},
\end{equation*}
where $\B a$ is the array steering vector for an angle of departure (arrival) $\theta_T$ ($\theta_R$) and the
power density $g$ is a sum of weighted Laplace densities where the standard deviation describes the angular spread \cite{3GPP}. For the case of a \ac{ula} at the \ac{BS}, it holds that $	\B a(\theta_R) =[1,\exp(j\pi\sin \theta_{R}),\dots,\exp(j(S-1) \pi\sin \theta_R)]\h$.
%\begin{equation*}
%	\B a(\theta_R) =[1,\exp(j\pi\sin \theta_{R}),\dots,\exp(j(S-1) \pi\sin \theta_R)]\h. 
%\end{equation*}
The same structure holds also for the \ac{MS} array vector.
Consequently, both the transmit and receive side covariance matrices are Toeplitz matrices, which makes the matrix $\Cdel$ a block-Toeplitz matrix with Toeplitz blocks.
The \ac{snr} is $\tr(\Cdel\B X\h\B X) / (SN\sigma^2)$.
%\begin{align}\label{eq:snr}
%\frac{\op E [ \B h\h\B X\h\B X\B h]}{\op E [\B z\h\B z]} =  \frac{\tr(\Cdel\B X\h\B X)}{\tr(\sigma^2\eye_{SN})} =
%	\frac{\tr(\Cdel\B X\h\B X)}{SN\sigma^2}.
%	\end{align}

\section{Derivation of the Proposed Estimator}
To derive the \ac{cnn} channel estimator for MIMO cases, we start with the definitions of the \ac{mmse} estimator similar to \cite{8272484} for SIMO cases.
Assuming knowledge of the prior parameters $\bdelt$, the conditional \ac{mmse} estimate of $\B h$ from $\B y$ would read as
\begin{align}
	\op E[\B h|\B y,\bdelt] &= \op E[\B h\B y\h|\bdelt]\op E[\B y\B y\h|\bdelt]\inv \B y\\
	&=\Cdel\B X\h(\B X\Cdel\B X\h + \sigma^2\eye_{SN})\inv\B y \label{eq:cond_mmse}
	= \Wdel\B y,
\end{align} 
which depends linearly on the observations. However, as the parameters $\bdelt$ are unknown in general, the law of total expectation is used to compute \cite{8272484}
\begin{align}
	\hat{\B h} &= \op E[\B h|\B y] = \op E[\op E[\B h|\B y,\bdelt]] 
	= \op E[\Wdel\B y|\B y] %= \op E[\Wdel|\B y] \B y
	= \Wha_\star(\B y)\B y.
\end{align} 
Thus, to obtain $\hat{\B h}$, the MMSE estimate $\Wha_\star$ of the \ac{mmse} 
filter $\Wdel$ has to be determined, which nonlineraly depends on $\B y$.

Bayes' theorem is used to state the \ac{mmse} filter as \cite{8272484}
\begin{equation}
	\label{eq:Wstar}
	\Wha_\star = \int p(\bdelt|\B y) \Wdel \op d\bdelt = \frac{\int p(\B y|\bdelt)\Wdel p(\bdelt)\op d\bdelt }{\int p(\B y|\bdelt) p(\bdelt)\op d\bdelt }.
\end{equation}
As shown in Appendix \ref{app:lemma1}, the \ac{mmse} filter is written~as
\begin{equation}\label{eq:mmse}
	\Wha(\Chat) = \frac{\int \exp(\tr(\B X\Wdel\Chat) + \log|\eye-\B X\Wdel|)\Wdel p(\bdelt)\op d\bdelt}{\int \exp(\tr(\B X\Wdel\Chat) + \log|\eye-\B X\Wdel|) p(\bdelt)\op d \bdelt},
\end{equation}	
with the scaled sample covariance matrix $	\Chat =\siginv\B y\B y\h$ as input.
The \ac{mmse} filter $\Wha_\star$ depends on the observations through $\Chat$ and is thus a nonlinear filter. For an arbitrary distribution $p(\bdelt)$, the \ac{mmse} filter in \eqref{eq:mmse} is not computable. To overcome this, we introduce the following assumption equivalent to \cite{8272484}.

\begin{assumption}
	The distribution $p(\bdelt)$ is discrete and uniform, i.e., we have a grid $\{\bdelt_i:i=1,\dots,P \}$ and $p(\bdelt_i) = 1 / P$.
\end{assumption}
With this assumption, the \ac{mmse} estimator is~evaluated~as
\begin{equation}\label{eq:ge}
	\Wha\GE(\Chat)=  \frac{1/P\sum_{i=1}^P \exp(\tr(\B X\B W\deli\Chat) +b_i)\B W\deli}{1/P\sum_{i=1}^P \exp(\tr(\B X\B W\deli\Chat) + b_i)},
\end{equation}
with $b_i = \log|\eye-\B X\B W\deli|$. We refer to this as the \ac{ge},
% After column-wise vectorization we get 
%\begin{equation}
%	\vect(\Wha\GE(\Chat))=  \B A^{(2)}\GE \frac{\exp(\B A^{(1),\op H}\GE\vect(\Chat) + \B b  )}{\B 1\T \exp(\B A^{(1),\op H}\GE\vect(\Chat) + \B b  )}
%\end{equation}	
%with 
%\begin{align}
%	\B A^{(1)}\GE &= [\vect(\B W_{\bdelt_1}\h\B X\h ),\dots,\vect(\B W_{\bdelt_P}\h\B X\h )],\\
%	\B A^{(2)}\GE &= [\vect(\B W_{\bdelt_1} ),\dots,\vect(\B W_{\bdelt_P} )],~~\B b = [b_1,\dots,b_P]\T
%\end{align}
which neglects the true continuous distribution of $\bdelt$, resulting in an approximation error that decreases with increasing number of samples $P$ \cite{8272484}. 
The order of complexity for evaluating the \ac{ge} is $\mathcal{O}(S^2U^2 P)$ in the case $U=N$ due to the matrix-matrix computations in \eqref{eq:ge}.
%However, with increasing number $P$ of samples the approximation error decreases, which also increases the complexity~of~the~estimator.

In the following we reduce the complexity of the estimator where we exploit common structure of covariance matrices.
%which is present for certain array geometries.
\begin{assumption}\label{ass:2}
	The filters $\B W\deli$ can be decomposed as 
	\begin{equation}\label{eq:ass2_se}
		\B W\deli = \B Q\h\diagblock_{S,U}(\B w\deli)\B Q\B X\h,
	\end{equation}
	with $\B Q\in\mathbb{C}^{SU\times SU}$ being the Kronecker product of two \ac{dft} matrices of size $S\times S$ and $U\times U$,~i.e.,
	\begin{equation}\label{eq:dft}
		\B Q = \B F_U\otimes \B F_S,
	\end{equation}
	and with the $\diagblock$ operator as defined in Appendix \ref{app:diagblock}. 
\end{assumption}
For further verification of Assumption \ref{ass:2}, see Appendix \ref{app:SE}.
With this assumption, the \ac{mmse} filter from \eqref{eq:ge} is approximately
\begin{align}\label{eq:ass2}
	\Wha(\cbar) \approx \B Q\h\diagblock_{S,U}\left(\B w(\cbar)
	\right)\B Q\B X\h ,
\end{align}
with the definitions
\begin{align}
	\B w(\cbar) &= \B A\frac{\exp(\B A\T\cbar + \B b)}{\B 1_P\T\exp(\B A\T\cbar + \B b)}
	= \B A \phi(\B A\T\cbar + \B b)
	,\label{eq:SE_element}\\
	\B A &=[\B w_{\bdelt_1}, \dots, \B w_{\bdelt_P}]\in	\mathbb{C}^{SU^2\times P}, \label{eq:ASE}\\
	\cbar&= \sigma^{-2}\diagblock_{S,U}(\B Q\B X\h\B y\B y\h\B X\B Q\h)\in \mathbb{C}^{SU^2} ,\label{eq:chat}
\end{align}
where we identify the softmax function $\phi$ in \eqref{eq:SE_element} and $\B b = [b_{\bdelt_1},...,b_{\bdelt_P}]\T$ as shown in Appendix \ref{app:SE}.

%\begin{equation}\label{eq:ass2}
%	\hat{\B h} = \B Q\h\diagblock_{S,U}(\B w\SE(\cbar))\B Q\B y.
%\end{equation} 
We further reduce the complexity by assuming $\B A$ to be a block-circulant matrix with circulant blocks, which is valid for specific scenarios, i.e., a single propagation cluster.
%In order to reasonably assume block-circulant matrices $\B A\SE$ in \eqref{eq:ASE}, several assumptions are needed. 
In \cite[Appendix D]{8272484}, a shift invariance of the power density is discussed in the SIMO case with a \ac{ula} at the \ac{BS}, which generalizes to a 2D shift invariance in MIMO cases. Thus, the matrix $\B A$ from \eqref{eq:ASE} consists of vectors $\B w\deli$ which are invariant to a certain vertical and horizontal shift, resulting in the block-circulant structure and motivating the following~assumption.
%In the MIMO cashalbe spalte noche, the covariance matrix can be decomposed as the Kronecker prodcut of the transmit and receive side covariance matrices.
%Therefore, the 1D shift invariance from \cite{8272484} for the single propagation path case in the 3GPP model extends to a 2D shift invariance and the matrix $\B A\SE$ can be approximated with the following assumption.
%\begin{equation}
%	\B A\SE = \B Q \diag(\B Q \B w_0)\B Q.
%\end{equation}
%This is equal to constructing $\B A\SE$ by a 2D circular convolution with the uniform samples $\B w_0$ as convolution kernel.
\begin{assumption}\label{ass:3}
	The matrix $\B A$ is block-circulant and therefore it exists a $\B w_0\in\mathbb{R}^{SU}$ such that 
	\begin{equation}
		\B A = \B Q\h\diag(\B Q\B w_0)\B Q,
	\end{equation}
	with $\B Q$ as defined in \eqref{eq:dft}.
\end{assumption}

This is equal to constructing $\B A$ by a 2D circular convolution with $\B w_0$ as convolution kernel.
Given the relationship between block-circulant matrices and 2D circular convolution, we write 
\begin{equation}
\B A \B x = \B Q\h \diag(\B Q\B w_0) \B Q \B x = \B w_0 \star \B x.\label{eq:2dcirc}
\end{equation}
%\begin{equation}
%	\B A \B x = \B Q\h \diag(\B Q\B a) \B Q \B x = \B a \star \B x.
%	\label{eq:2d_conv}
%\end{equation}
Here, we compute the 2D convolution of the two vectorized quantities $\B w_0$ and $\B x$.
Hence, in a first step, a reshaping of both $\B w_0$ and $\B x$ into matrices with $S$ rows is necessary.
Afterwards, we vectorize the result, which yields the vector $\B A\B x$.
This three-step convolution process~is~denoted~by~$\B w_0\star \B x$~for~simplicity.

%Here and in the following, we define for the ease of notation the 2D convolution operator $\B x\star\B y$ for two vectors $\B x$, $\B y$ such that it combines the reshaping into matrices with $S$ rows before and the column-wise vectorization after the 2D convolution. This reshaping has to be done as the 2D convolution is only reasonable for 2D inputs, but we work with the vectorized model.
%More details why Assumption \ref{ass:3} is reasonable for a certain scenario is given in Appendix \ref{app:ass3}. 
%We refer to the estimator incorporating Assumption \ref{ass:3} as \ac{fe}.

\section{Orthogonal Pilot Sequences}
The general formulation of the estimator breaks down to a less complex implementation if we use pilot matrices with certain properties. More precisely, we choose $\B X^\prime =\frac{1}{\sqrt{U}}\B F_{U\times N}$
%\begin{equation}
%	\B X^\prime =U^{-\frac{1}{2}} \B F_{U\times N}
%\end{equation}
where $\B F_{U\times N}$ contains the first $U$ rows of a $N\times N$ \ac{dft} matrix for the case $U\leq N$.
This ensures $\B X\h\B X = \frac{N}{U}\eye_{SU}$, which is also endorsed in \cite{1597555}, \cite{1045284}.
With this, the matrix product $\tilde{\B Q}\tilde{\B Q}\h$ in \eqref{eq:se_derivation} becomes diagonal and Assumption 2 can be written in terms of $\B W\deli = \B Q\h \diag(\B w\deli)\B Q\B X\h$.
%\begin{equation}\label{eq:se_diag}
%	\B W\deli = \B Q\h \diag(\B w\deli)\B Q\B X\h.
%\end{equation}
Also, the estimator input in \eqref{eq:chat} simplifies to $\chat = \sigma^{-2}|\B Q\B X\h\B y|^2$.
%The estimator incorporating Assumptions \ref{ass:1} and \ref{ass:2} is then given 
%\begin{equation}
%	\Wha\SE(\hat{\B c}) =
%	\B Q\h\operatorname{diag}\left(\B A\SE\frac{\exp(\B A\SE\T\chat + \B b)}{\B 1_P\T\exp(\B A\SE\T\chat + \B b)}
%	\right)\B Q\B X\h 
%\end{equation}
%with the matrix
%\begin{equation}
%	\mathbb{C}^{SU\times P}\ni\B A\SE = \begin{bmatrix}
%		\B w_{\bdelt_1}&\cdots &\B w_{\bdelt_P}
%	\end{bmatrix},
%\end{equation}
%containing the element-wise \ac{mmse} filters
%and the pre-processed observations $	\chat = \sigma^{-2}|\B Q\B X\h\B y|^2$.
%%\begin{equation}
%%	\chat = \siginv|\B Q\B X\h\B y|^2.
%%\end{equation}
%Note that the weight matrix $\B A\SE$ and the input $\chat$ have by a factor $U$ less entries than in the general case in \eqref{eq:ASE}.
In extension to the property in \eqref{eq:2dcirc}, due to the 2D shift-invariance and the circular assumption, the matrix-vector product $\B A\T\chat$ can be written as a 2D circular~convolution, i.e.,
\begin{align}
	\B A\T\chat = \tilde{\B w_0} \star \B \chat,\label{eq:circ}
%		\begin{aligned}
%	\B A\chat = \B Q\h\diag(\B Q \B w_0)\B Q\chat = \B w_0 \star \B \chat,
%	\label{eq:circ}
%	\\
%	\B A\T\chat = \B Q\h\diag(\B Q\tilde{ \B w}_0)\B Q\chat = \tilde{\B w_0} \star \B \chat,
%	\label{eq:circ2}
	%	\end{aligned}
\end{align}
where $\tilde{\B w}_0$ contains the entries of $\B w_0$ in reversed order.
We refer to the estimator containing all three assumptions as the \ac{fe}, which is given by
\begin{equation}\label{eq:fe}
	\Wha(\hat{\B c}) =
	\B Q\h\operatorname{diag}\left(\B w_0 \star \phi(\tilde{\B w}_0\star\chat+ \B b)
	\right)\B Q\B X\h ,
\end{equation}
where the $\diag(\cdot)$ operator replaces the $\diagblock(\cdot)$ operator from \eqref{eq:ass2} due to the orthogonal pilot sequences. The 2D circular convolution is justified due to the properties from \eqref{eq:circ}.
Note that the \ac{fe} is also applicable in the case of arbitrary pilot sequences, but with higher complexity.

The \ac{fe} has low complexity, but the underlying assumptions only hold for many antennas and a single propagation cluster which may be rarely fulfilled in real scenarios. Therefore, we interpret the estimator from \eqref{eq:fe} as a \ac{cnn} with two 2D convolution layers, which implements a function from the set
\begin{equation}
	\mathcal{W}_{\text{CNN}} = \{\B x \mapsto \B a^{(2)} \star \psi (\B a^{(1)} \star \B x + \B b^{(1)}) + \B b^{(2)}
	\},
\end{equation}
where $\B a^{(l)}, \B b^{(l)}\in\mathbb{R}^{SU} , l=1,2$ are the parameters which are learned during training from samples $(\B y_i, \B h_i)$, generated by the 3GPP model.
Although we identified the softmax function $\phi$ as the activation in \eqref{eq:SE_element}, we relax the activation to be a different function $\psi$, e.g. the well-known \ac{relu}. This can lead to a better generalization and convergence ability.
Note that in contrast to common linear convolution layers in \acp{cnn}, the above derivation suggests to use circular convolution layers in the proposed \ac{cnn}.
The optimal \ac{cnn} estimator is the function which minimizes the MSE, i.e.,
\begin{equation}
	\hat{\B w}_{\text{CNN}} = \underset{ \hat{\B w}(\cdot)\in\mathcal{W}_{\text{CNN}}}{\arg\min}~\op E[||\B h - \B Q\h\operatorname{diag}(\hat{\B w}(\chat))\B Q\B X\h \B y||_2^2].
\end{equation}

The complexity of the \ac{cnn} estimator is equal to that of the \ac{fe} as learning is done offline. Due to the \ac{fft} for the 2D circular convolution and the pilot matrix, the complexity of the estimator is only $\mathcal{O}(SU\log(SU))$ which is a drastic decrease compared to the \ac{ge} complexity of $\mathcal{O}(S^2U^2 P)$.
%\begin{equation}
%	\mathcal{O}(SK\log(SK))
%\end{equation}

%\section{Reference Algorithms}

%\clearpage
\section{Simulation Results}
\begin{figure}[t]
%	\centering
	\begin{subfigure}[t]{1\columnwidth}
		\begin{tikzpicture}
%			\pgfplotsset{
%				compat=1.3,
%				tick label style={font=\scriptsize},
%				label style={font=\scriptsize},
%				legend style={font=\scriptsize}	
%			}
			\begin{axis}
				[width=1\columnwidth,
				height=0.5\columnwidth,
				xtick=data, 
				xmin=-15, 
				xmax=20,
				xlabel={SNR $[\operatorname{dB}]$},
				ymode = log, 
				ymin=3*1e-3,
				ymax=1,
				ylabel= {Normalized MSE}, 
				ylabel shift = 0.0cm,
				grid = both,
				legend columns = 2,
				legend entries={
					genie,
					GE,
					ML,
					%					SE,
					FE,
					softmax,
					ReLU,
					LS,
					genie-OMP,
				},
				legend style={at={(0.0,0.0)}, anchor=south west},
				]
				
				\addplot[mark options={solid},color=TUMBlack,line width=1.2pt]
				table[x= snr, y=nmse, col sep=comma]
				{csvdat/genieMMSE_MIMO/n_paths_3_angle_spreadBS_2_angle_spreadMS_35_n_antennasBS_64_n_antennasMS_2_snr_min_-15.0_snr_max_20_unquant_2pilot.csv};
				
				\addplot[mark options={solid},color=TUMBlack,line width=1.2pt,dashed]
				table[x= SNR, y=GE, col sep=comma]
				{csvdat/MIMO/GE/2020-11-05_08-48-33_3paths_64BS_antennas_2MS_antennas_35AS_20lbatch_400ebatch_2pilots.csv};

				\addplot[mark options={solid},color=TUMBeamerDarkRed,line width=1.2pt,dash dot,mark=o]
				table[x= SNR, y=ML_fft, col sep=comma]
				{csvdat/MIMO/ML/2020-11-04_21-15-15_3paths_64BS_antennas_2MS_antennas_35AS_20lbatch_400ebatch_2pilots.csv};

				%				\addplot[mark options={solid},color=TUMBeamerDarkRed,line width=1pt,mark = diamond,dashed]
				%				table[x= SNR, y=SE_fft, col sep=comma]
				%				{csvdat/MIMO/SE/2020-11-06_07-44-28_3paths_64BS_antennas_2MS_antennas_35AS_20lbatch_400ebatch_2pilots.csv};
				
				\addplot[mark options={solid},color=TUMBeamerGreen,line width=1.2pt,mark = triangle,dashed]
				table[x= SNR, y=FE, col sep=comma]
				{csvdat/MIMO/FE/2020-11-06_12-47-18_3paths_64BS_antennas_2MS_antennas_35AS_20lbatch_400ebatch_2pilots.csv};
				
				\addplot[mark options={solid},color=TUMBeamerOrange,line width=1.2pt,mark = triangle]
				table[x= SNR, y=cnn_fft_softmax_non_hier_False, col sep=comma]
				{csvdat/MIMO/CNN/2020-11-05_22-46-01_3paths_64BS_antennas_2MS_antennas_35AS_20lbatch_400ebatch_2pilots.csv};

				\addplot[mark options={solid},color=TUMBlue,line width=1.2pt,mark = o]
				table[x= SNR, y=cnn_fft_relu_non_hier_False, col sep=comma]
				{csvdat/MIMO/CNN/2020-11-04_20-54-43_3paths_64BS_antennas_2MS_antennas_35AS_20lbatch_400ebatch_2pilots.csv};
				
				\addplot[mark options={solid},color=TUMBlack,line width=1.2pt,mark = square,dotted]
				table[x= SNR, y=LS, col sep=comma]
				{csvdat/LS/2020-11-06_07-24-01_3paths_64BS_antennas_2MS_antennas_35AS_20lbatch_20ebatch_2pilots.csv};
				
				\addplot[mark options={solid},color=TUMBlack,line width=1.2pt,mark = x,dotted]
				table[x= SNR, y=Genie_OMP, col sep=comma]
				{csvdat/MIMO/OMP/2020-11-09_15-57-44_3paths_64BS_antennas_2MS_antennas_35AS_20lbatch_400ebatch_2pilots.csv};
				
			\end{axis}
		\end{tikzpicture}
	\end{subfigure}
	\begin{subfigure}[t]{0.51\columnwidth}
%		\raggedleft
		\begin{tikzpicture}
%			\pgfplotsset{
%				compat=1.3,
%				tick label style={font=\scriptsize},
%				label style={font=\scriptsize},
%				legend style={font=\tiny}	
%			}
			\begin{axis}
				[width=1\columnwidth,
				height=1\columnwidth,
				xtick=data, 
				xmin=-15, 
				xmax=20,
				xlabel={SNR $[\operatorname{dB}]$},
				ymode = log, 
				ymin=1e-3,
				ymax=1,
				ylabel= { Normalized MSE}, 
				ylabel shift =-0.1cm,
				grid = both,
				legend columns = 2,
				legend style={nodes={scale=0.5}},
				legend entries={
					%			\scriptsize  genie,
					%			\scriptsize GE,
					%			\scriptsize SE,
					%			\scriptsize ML,
					%			\scriptsize FE,
					%			\scriptsize softmax,
					%			\scriptsize ReLU,
					%			\scriptsize LS,
					%			\scriptsize genie-OMP,
				},
				legend style={at={(0.0,0.0)}, anchor=south west},
				]

				\addplot[mark options={solid},color=TUMBlack,line width=1.2pt]
				table[x= snr, y=nmse, col sep=comma]
				{csvdat/genieMMSE_MIMO/n_paths_1_angle_spreadBS_2_angle_spreadMS_35_n_antennasBS_64_n_antennasMS_2_snr_min_-15.0_snr_max_20_unquant_2pilot.csv};
				
				\addplot[mark options={solid},color=TUMBlack,line width=1.2pt,dashed]
				table[x= SNR, y=GE, col sep=comma]
				{csvdat/MIMO/GE/2020-11-05_08-47-10_1paths_64BS_antennas_2MS_antennas_35AS_20lbatch_400ebatch_2pilots.csv};

				%				\addplot[mark options={solid},color=TUMBeamerDarkRed,line width=1pt,mark = diamond,dashed]
				%				table[x= SNR, y=SE_fft, col sep=comma]
				%				{csvdat/MIMO/SE/2020-10-24_11-22-59_1paths_64BS_antennas_2MS_antennas_35AS_20lbatch_200ebatch_2pilots.csv};
				
				\addplot[mark options={solid},color=TUMBeamerDarkRed,line width=1.2pt,mark = o,dash dot]
				table[x= SNR, y=ML_fft, col sep=comma]
				{csvdat/MIMO/ML/2020-11-04_21-16-09_1paths_64BS_antennas_2MS_antennas_35AS_20lbatch_400ebatch_2pilots.csv};
				
				\addplot[mark options={solid},color=TUMBeamerGreen,line width=1.2pt,mark = triangle,dashed]
				table[x= SNR, y=FE, col sep=comma]
				{csvdat/MIMO/FE/2020-11-06_12-46-13_1paths_64BS_antennas_2MS_antennas_35AS_20lbatch_400ebatch_2pilots.csv};
				
				\addplot[mark options={solid},color=TUMBeamerOrange,line width=1.2pt,mark = triangle]
				table[x= SNR, y=cnn_fft_softmax_non_hier_False, col sep=comma]
				{csvdat/MIMO/CNN/2020-11-05_22-48-16_1paths_64BS_antennas_2MS_antennas_35AS_20lbatch_400ebatch_2pilots.csv};
				
				\addplot[mark options={solid},color=TUMBlue,line width=1.2pt,mark = o]
				table[x= SNR, y=cnn_fft_relu_non_hier_False, col sep=comma]
				{csvdat/MIMO/CNN/2020-11-04_21-02-36_1paths_64BS_antennas_2MS_antennas_35AS_20lbatch_400ebatch_2pilots.csv};
				
				\addplot[mark options={solid},color=TUMBlack,line width=1.2pt,mark = square,dotted]
				table[x= SNR, y=LS, col sep=comma]
				{csvdat/LS/2020-11-06_07-19-35_1paths_64BS_antennas_2MS_antennas_35AS_20lbatch_20ebatch_2pilots.csv};
				
				\addplot[mark options={solid},color=TUMBlack,line width=1.2pt,mark = x,dotted]
				table[x= SNR, y=Genie_OMP, col sep=comma]
				{csvdat/MIMO/OMP/2020-11-09_15-54-37_1paths_64BS_antennas_2MS_antennas_35AS_20lbatch_400ebatch_2pilots.csv};
				
				%			\addplot[mark options={solid},color=TUMBlue,line width=1.2pt,mark = o,dashed]
				%			table[x= SNR, y=cnn_fft_relu_non_hier_False, col sep=comma]
				%			{csvdat/MIMO/CNN/2020-11-09_08-51-12_1paths_64BS_antennas_2MS_antennas_35AS_20lbatch_400ebatch_2pilots.csv};
				
			\end{axis}
		\end{tikzpicture}
	\end{subfigure}%
	\begin{subfigure}[t]{0.51\columnwidth}
%		\raggedright
		\begin{tikzpicture}
%			\hspace{-0.15cm}
%			\pgfplotsset{
%				compat=1.3,
%				tick label style={font=\scriptsize},
%				label style={font=\scriptsize},
%			}
			\begin{axis}
				[width=1\columnwidth,
				height=1\columnwidth,
				xtick=data, 
				xmin=-15, 
				xmax=20,
				xlabel={SNR $[\operatorname{dB}]$},
				ymode = log, 
				ymin=6*1e-3,
				ymax=1,
				ylabel shift = 0cm,
				grid = both,
				legend columns = 2,
				legend entries={
					%			\scriptsize  genie,
					%			\scriptsize GE,
					%			\scriptsize SE,
					%			\scriptsize ML,
					%			\scriptsize FE,
					%			\scriptsize softmax,
					%			\scriptsize ReLU,
					%			\scriptsize LS,
					%			\scriptsize genie-OMP,
				},
				legend style={at={(0.0,0.0)}, anchor=south west},
				]
				
				\addplot[mark options={solid},color=TUMBlack,line width=1.2pt]
				table[x= snr, y=nmse, col sep=comma]
				{csvdat/genieMMSE_MIMO/n_paths_10_angle_spreadBS_2_angle_spreadMS_35_n_antennasBS_64_n_antennasMS_2_snr_min_-15.0_snr_max_20.0_unquant_2pilot.csv};
				
				\addplot[mark options={solid},color=TUMBlack,line width=1.2pt,dashed]
				table[x= SNR, y=GE, col sep=comma]
				{csvdat/MIMO/10paths/2020-11-28_12-15-05_10paths_64BS_antennas_2MS_antennas_35AS_40lbatch_100ebatch_10lbatchsize_2pilots_2trainpilots.csv};
				
				%				\addplot[mark options={solid},color=TUMBeamerDarkRed,line width=1pt,mark = diamond,dashed]
				%				table[x= SNR, y=SE_fft, col sep=comma]
				%				{csvdat/MIMO/10paths/2020-11-28_10-32-05_10paths_64BS_antennas_2MS_antennas_35AS_40lbatch_400ebatch_10lbatchsize_2pilots_[2]trainpilots.csv};
				
				\addplot[mark options={solid},color=TUMBeamerDarkRed,line width=1.2pt,dash dot,mark=o]
				table[x= SNR, y=ML_fft, col sep=comma]
				{csvdat/MIMO/10paths/2020-11-28_10-32-05_10paths_64BS_antennas_2MS_antennas_35AS_40lbatch_400ebatch_10lbatchsize_2pilots_2trainpilots.csv};
				
				\addplot[mark options={solid},color=TUMBeamerGreen,line width=1.2pt,mark = triangle,dashed]
				table[x= SNR, y=FE, col sep=comma]
				{csvdat/MIMO/10paths/2020-11-28_10-32-05_10paths_64BS_antennas_2MS_antennas_35AS_40lbatch_400ebatch_10lbatchsize_2pilots_2trainpilots.csv};
				
				\addplot[mark options={solid},color=TUMBeamerOrange,line width=1.2pt,mark = triangle]
				table[x= SNR, y=cnn_fft_softmax_non_hier_False, col sep=comma]
				{csvdat/MIMO/10paths/2020-12-02_11-21-36_10paths_64BS_antennas_2MS_antennas_35AS_40lbatch_400ebatch_20lbatchsize_2pilots_2trainpilots.csv}; 

				\addplot[mark options={solid},color=TUMBlue,line width=1.2pt,mark = o]
				table[x= SNR, y=cnn_fft_relu_non_hier_False, col sep=comma]
				{csvdat/MIMO/10paths/2020-11-28_15-23-43_10paths_64BS_antennas_2MS_antennas_35AS_40lbatch_400ebatch_50lbatchsize_2pilots_2trainpilots.csv};
				\addplot[mark options={solid},color=TUMBlack,line width=1.2pt,mark = square,dotted]
				table[x= SNR, y=Genie_Aided, col sep=comma]
				{csvdat/MIMO/10paths/2020-11-28_10-39-51_10paths_64BS_antennas_2MS_antennas_35AS_40lbatch_400ebatch_10lbatchsize_2pilots_2trainpilots.csv};
				
				\addplot[mark options={solid},color=TUMBlack,line width=1.2pt,mark = x,dotted]
				table[x= SNR, y=Genie_OMP, col sep=comma]
				{csvdat/MIMO/10paths/2020-11-28_10-39-51_10paths_64BS_antennas_2MS_antennas_35AS_40lbatch_400ebatch_10lbatchsize_2pilots_2trainpilots.csv};

			\end{axis}
		\end{tikzpicture}
	\end{subfigure}
	\caption{Performance of estimators with $S=64$ and $N=U=2$. Top: 3 clusters, Bottom: 1 cluster (left), 10 clusters (right).}
	\label{fig:snr_diff_paths}
	\vspace{-0.5cm}
\end{figure}
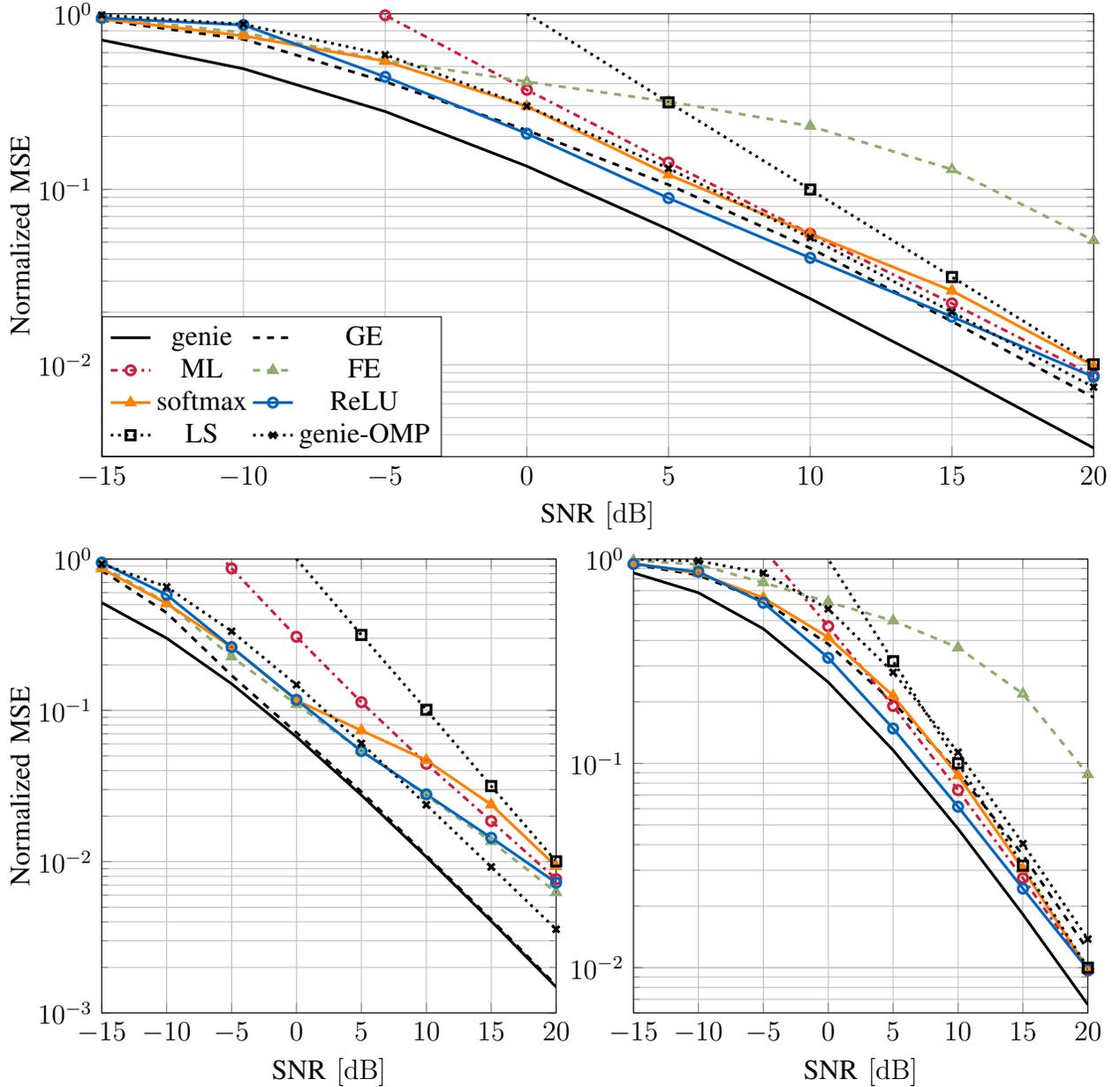
We consider the 3GPP channel model \cite{3GPP} with parameters $\bdelt$ drawn uniformly at random for each channel realization, which means that a different covariance matrix $\Cdel$ applies. We choose the per-sample noise variance $\sigma^2$ by the given \ac{snr},
and we assume that the prior $\bdelt$ is known during training.
As performance measure we use the MSE, normalized by $SU$.
%\begin{equation}
%	\frac{\op E[||\B H - \hat{\B H}||_F^2]}{SU} = \frac{\op E[||\B h - \hat{\B h}||_2^2]}{SU}.
%\end{equation}

In the 3GPP channel model, the covariance matrices have a low numerical rank \cite{7499112}, and therefore, the channel matrix can be approximated by a row-sparse matrix $\B T$ for a given dictionary $\B D$, i.e.,
$\B h \approx (\eye_U \otimes \B D)\B t$ with $\B t = \vect(\B T)$.
%Vectorizing this expression results in 
%$\B h \approx (\eye_M \otimes \B D)\B t$
%with $\B t = \vect(\B T)$.
A reasonable choice for the dictionary $\B D$ is an oversampled \ac{dft} matrix \cite{7499112}. 
The vector $\B t$ can then be found by solving a sparse approximation problem, for which the \ac{omp} algorithm is suitable \cite{681706}.
%\begin{equation}
%	\B t = \underset{\B t:|\operatorname{supp}(\B t)|\leq k}{\arg\min}~~~||\B y - \B X(\eye_M\otimes \B D)\B t||^2,
%\end{equation}
%where $|\operatorname{supp}(\B t)|$ denotes the cardinality of the support $\operatorname{supp}(\B t)$ of $\B t$.
%A well-known approach to solve this problem is the \ac{omp} algorithm \cite{681706}.
As the sparsity order is unknown, we use a genie-aided upper bound in our simulations, which decides about the sparsity level with the given exact channel realization. 
%This is clearly an upper bound for the performance of \ac{omp}.

Another low-complexity \ac{ce} algorithm is \ac{ml} estimation of the structured covariance matrix \cite{7051818}, \cite{1456695}. The estimate with the block-circular assumption for the covariance matrix is
%\begin{equation}
%	\hat{\B h} = \B Q\h\sigma^{-2}(\sigma^{-2}\tilde{\B Q}\tilde{\B Q}\h + \diag(\B c_{\bdelt}^{\text{ML}})\inv)\inv\tilde{\B Q}\B y
%\end{equation}
\begin{equation*}
	\hat{\B h} = \B Q\h\diag(\B c_{\bdelt}^{\text{ML}})(NU\inv\diag(\B c_{\bdelt}^{\text{ML}}) + \sigma^2\eye_{SU})\inv\tilde{\B Q}\B y
\end{equation*}
with $\tilde{\B Q} = \B Q\B X\h$. 
The eigenvalues of $\Cdel$ are estimated as
$\B c_{\bdelt}^{\text{ML}} = [\B s - \sigma^2\B 1]_+$, where $\B s = |\B Q \B X\h \B y|^2$ and the $i$th element of $[\B x]_+$ is $\max( x_i,0)$, cf. \cite{8272484}.
%The estimated power spectrum is $\B s = |\B Q \B X\h \B y|^2$,
%and the eigenvalues of $\Cdel$ are estimated as
%\begin{equation}
%	\B c_{\bdelt}^{\text{ML}} = [\B s - \sigma^2\B 1]_+
%\end{equation}
%$\B c_{\bdelt}^{\text{ML}} = [\B s - \sigma^2\B 1]_+$, where the $i$th element of $[\B x]_+$ is $\max( x_i,0)$, cf. \cite{8272484}.
We further show the \ac{ls} solution which minimizes the $\ell_2$ norm $||\B y - \B X\B h||_2$ \cite{1597555}.

We compare the \ac{cnn} estimators with softmax and \ac{relu} activation with the reference algorithms described above. Further, we show the derived \ac{ge} for $P$=$16S$ samples and the \ac{fe}. As a utopian lower bound we show the genie-aided \ac{mmse} \eqref{eq:cond_mmse}, which has perfect knowledge of the channel covariance matrix $\Cdel$ for each channel realization.
The training procedure of the proposed estimator is as follows. We train the described architecture for 250 epochs consisting of 40 batches with a batch-size of 20. We initialize the weights randomly according to a truncated normal distribution with a certain variance. We use the Adam optimizer \cite{adam} and $\ell_2$ regularization. More informations can be found in \cite{sim_code_new}.

Fig. \ref{fig:snr_diff_paths} shows the NMSE vs. SNR for a setup with $S$=$64$ and $U$=$N$=2 for channels with one, three and ten propagation clusters. The performance of the \ac{ge} which is based on Assumption 1 is close to the utopian genie-\ac{mmse}, but degrades for higher numbers of propagation clusters. 
The \ac{fe}, incorporating all three assumptions, is only reasonable for a single propagation cluster which can be clearly observed due to the high performance loss in scenarios with more clusters. 
This gives now room where the learning-based \ac{cnn} estimator which is based on the same low-complexity structure as the \ac{fe}, can shine. Especially, the approach with \ac{relu} activation is advantageous, showing strong perormance in all settings and being able to outperform all reference algorithms and the \ac{ge}, especially with an increasing number of clusters. The \ac{ml} approach only performs well in high \ac{snr} regions, whereas genie-\ac{omp} perfoms reasonable well over the whole \ac{snr} range, but with decreasing performance for higher numbers of clusters. The \ac{ls} estimator is only competitive for very high \ac{snr} values.

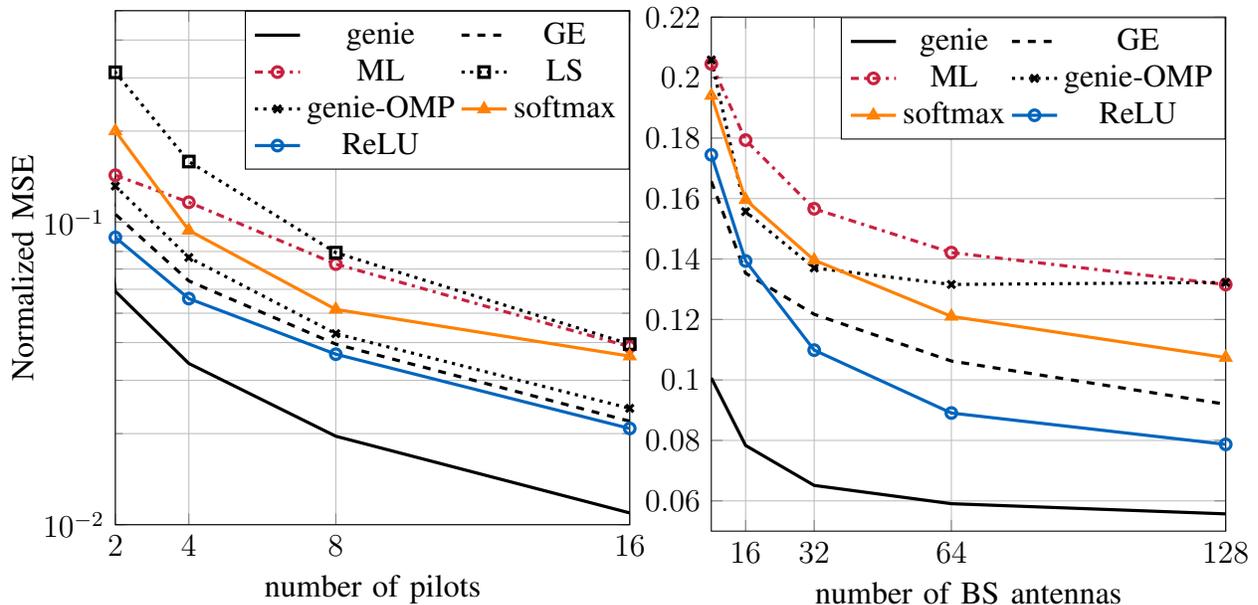
\begin{figure}[t]
	\centering
	\begin{subfigure}[t]{0.51\columnwidth}
		\begin{tikzpicture}
%			\pgfplotsset{
%				compat=1.3,
%				tick label style={font=\scriptsize},
%				label style={font=\scriptsize},
%				legend style={font=\scriptsize}	
%			}
			\begin{axis}
				[width=1\columnwidth,
				height=1\columnwidth,
				xtick=data, 
				xmin=2, 
				xmax=16,
				xlabel={number of pilots},
				ymode = log, 
				ymin=1e-2,
				ymax=0.5,
				ylabel= {Normalized MSE}, 
				ylabel shift = 0.0cm,
				grid = both,
				legend columns = 2,
				legend entries={
					genie,
					GE,
					%				 SE,
					ML,
					%				\footnotesize FE,
					%				\footnotesize ML,
					LS,
					genie-OMP,
					softmax,
					ReLU,
				},
				legend style={at={(1.0,1.0)}, anchor=north east},
				]
				
				\addplot[mark options={solid},color=TUMBlack,line width=1.2pt]
				table[x= pilots, y=genie, col sep=comma]
				{csvdat/MIMO/plot_pilots/2020-11-12_09-34-09_3paths_64BS_2MS_antennas_35AS_20lbatch_400ebatch.csv};
				
				\addplot[mark options={solid},color=TUMBlack,line width=1.2pt,dashed]
				table[x= pilots, y=GE, col sep=comma]
				{csvdat/MIMO/plot_pilots/2020-11-12_09-34-09_3paths_64BS_2MS_antennas_35AS_20lbatch_400ebatch.csv};
				%			%			
				%			\addplot[mark options={solid},color=TUMBeamerDarkRed,line width=1pt,mark = diamond,dashed]
				%			table[x= pilots, y=SE, col sep=comma]
				%			{csvdat/MIMO/plot_pilots/2020-11-12_09-34-09_3paths_64BS_2MS_antennas_35AS_20lbatch_400ebatch.csv};
				%				
				\addplot[mark options={solid},color=TUMBeamerDarkRed,line width=1.2pt,dash dot,mark=o]
				table[x= pilots, y=ML, col sep=comma]
				{csvdat/MIMO/ML/ML_pilots.csv};
				
				\addplot[mark options={solid},color=TUMBlack,line width=1.2pt,mark = square,dotted]
				table[x= pilots, y=LS, col sep=comma]
				{csvdat/MIMO/plot_pilots/2020-11-12_09-34-09_3paths_64BS_2MS_antennas_35AS_20lbatch_400ebatch.csv};
				\addplot[mark options={solid},color=TUMBlack,line width=1.2pt,mark = x,dotted]
				table[x= pilots, y=OMP, col sep=comma]
				{csvdat/MIMO/plot_pilots/2020-11-12_09-34-09_3paths_64BS_2MS_antennas_35AS_20lbatch_400ebatch.csv};
				\addplot[mark options={solid},color=TUMBeamerOrange,line width=1.2pt,mark = triangle]
				table[x= pilots, y=softmax, col sep=comma]
				{csvdat/MIMO/plot_pilots/2020-11-12_09-34-09_3paths_64BS_2MS_antennas_35AS_20lbatch_400ebatch.csv};
				
				\addplot[mark options={solid},color=TUMBlue,line width=1.2pt,mark = o]
				table[x= pilots, y=relu_hier, col sep=comma]
				{csvdat/MIMO/plot_pilots/2020-11-12_09-34-09_3paths_64BS_2MS_antennas_35AS_20lbatch_400ebatch.csv};
			\end{axis}
		\end{tikzpicture}
		%	\caption{Performance of different estimators with $S=64$, $U=2$, 3 paths, SNR=$5$\dB.}
		%	\label{fig:pilots}
		%\end{figure}
	\end{subfigure}%
	\begin{subfigure}[t]{0.51\columnwidth}
		%\begin{figure}[ht]
		%	\centering
		\begin{tikzpicture}
%			\pgfplotsset{
%				compat=1.3,
%				tick label style={font=\scriptsize},
%				label style={font=\scriptsize},
%				legend style={font=\scriptsize}	
%			}
			\begin{axis}
				[width=1\columnwidth,
				height=1\columnwidth,
				xtick={16,32,64,128}, 
				xmin=8, 
				xmax=128,
				xlabel={number of BS antennas},
				%			ymode = log, 
				ymin=5*1e-2,
				ymax=2.2*1e-1,
				%			ymax=3.5*1e-1,
				tick label style={/pgf/number format/fixed},
				ylabel shift = 0.0cm,
				grid = both,
				legend columns = 2,
				legend entries={
					genie,
					GE,
					%				 SE,
					%				\footnotesize FE,
					ML,
					%			     LS,
					genie-OMP,
					softmax,
					ReLU,
					%				\footnotesize relu w/o RS,
				},
				legend style={at={(1.0,1.0)}, anchor=north east},
				]
				
				\addplot[mark options={solid},color=TUMBlack,line width=1.2pt]
				table[x= BS_antennas, y=genie, col sep=comma]
				{csvdat/MIMO/plot_antenna_BS/2020-11-12_09-34-09_3paths_2MS_antennas_35AS_20lbatch_400ebatch.csv};
				
				\addplot[mark options={solid},color=TUMBlack,line width=1.2pt,dashed]
				table[x= BS_antennas, y=GE, col sep=comma]
				{csvdat/MIMO/plot_antenna_BS/2020-11-12_09-34-09_3paths_2MS_antennas_35AS_20lbatch_400ebatch.csv};
				%			%			
				%			\addplot[mark options={solid},color=TUMBeamerDarkRed,line width=1pt,mark = diamond,dashed]
				%			table[x= BS_antennas, y=SE, col sep=comma]
				%			{csvdat/MIMO/plot_antenna_BS/2020-11-12_09-34-09_3paths_2MS_antennas_35AS_20lbatch_400ebatch.csv};
				
				%			\addplot[mark options={solid},color=TUMBeamerGreen,line width=1pt,mark = triangle,dashed]
				%			table[x= BS_antennas, y=FE, col sep=comma]
				%			{csvdat/MIMO/plot_antenna_BS/2020-11-12_09-34-09_3paths_2MS_antennas_35AS_20lbatch_400ebatch.csv};
				
				\addplot[mark options={solid},color=TUMBeamerDarkRed,line width=1.2pt,dash dot,mark=o]
				table[x= BS_antennas, y=ML, col sep=comma]
				{csvdat/MIMO/plot_antenna_BS/2020-11-12_09-34-09_3paths_2MS_antennas_35AS_20lbatch_400ebatch.csv};
				
				%			\addplot[mark options={solid},color=TUMBlack,line width=1pt,mark = square,dotted]
				%			table[x= BS_antennas, y=LS, col sep=comma]
				%			{csvdat/MIMO/plot_antenna_BS/2020-11-12_09-34-09_3paths_2MS_antennas_35AS_20lbatch_400ebatch.csv};
				%			
				\addplot[mark options={solid},color=TUMBlack,line width=1.2pt,mark = x,dotted]
				table[x= BS_antennas, y=OMP, col sep=comma]
				{csvdat/MIMO/plot_antenna_BS/2020-11-12_09-34-09_3paths_2MS_antennas_35AS_20lbatch_400ebatch.csv};
				\addplot[mark options={solid},color=TUMBeamerOrange,line width=1.2pt,mark = triangle]
				table[x= BS_antennas, y=softmax, col sep=comma]
				{csvdat/MIMO/plot_antenna_BS/2020-11-12_09-34-09_3paths_2MS_antennas_35AS_20lbatch_400ebatch.csv};
				
				\addplot[mark options={solid},color=TUMBlue,line width=1.2pt,mark = o]
				table[x= BS_antennas, y=relu, col sep=comma]
				{csvdat/MIMO/plot_antenna_BS/2020-11-12_09-34-09_3paths_2MS_antennas_35AS_20lbatch_400ebatch.csv};
				
			\end{axis}
		\end{tikzpicture}
		%	\caption{Performance of different estimators with $S=64$, $U=2$, 3 paths, SNR=$5$\dB.}
		%	\caption{Performance of different estimators with $U=N=2$, 3 paths, SNR=$5$\dB.}
		%	\label{fig:antennas}
		%\end{figure}
	\end{subfigure}
	\caption{Performance of different estimators with $U=2$, 3 clusters, SNR=$5$\dB. Left: $S=64$, Right: $N=2$}
	\label{fig:pilots+antennas}
	\vspace{-0.5cm}
\end{figure}

Fig. \ref{fig:pilots+antennas} (left) shows the performance for different numbers of pilots for again $S$=$64$ and $U$=$2$ with three propagation clusters for a fixed \ac{snr} of $5$\dB. The genie-\ac{omp} performs well, being able to reach the \ac{ge} performance. However, the \ac{cnn} approach with \ac{relu} activation is again able to outperform the reference algorithms as well as the \ac{ge} for all numbers~of~pilots. 
Fig. \ref{fig:pilots+antennas} (right) depicts the performance for different numbers of \ac{BS} antennas for $U$=$N$=$2$ and three propagation clusters for a fixed \ac{snr} of $5$\dB. The proposed learning-based approaches benefit from higher numbers of antennas, which is why the performance gap to the reference algorithms, which tend to saturate, slightly increases for higher numbers of antennas. Also, the gap between the \ac{cnn} with \ac{relu} and the genie-\ac{mmse} decreases for higher numbers of \ac{BS} antennas. The \ac{ls} estimator has the worst performance and is not shown. 

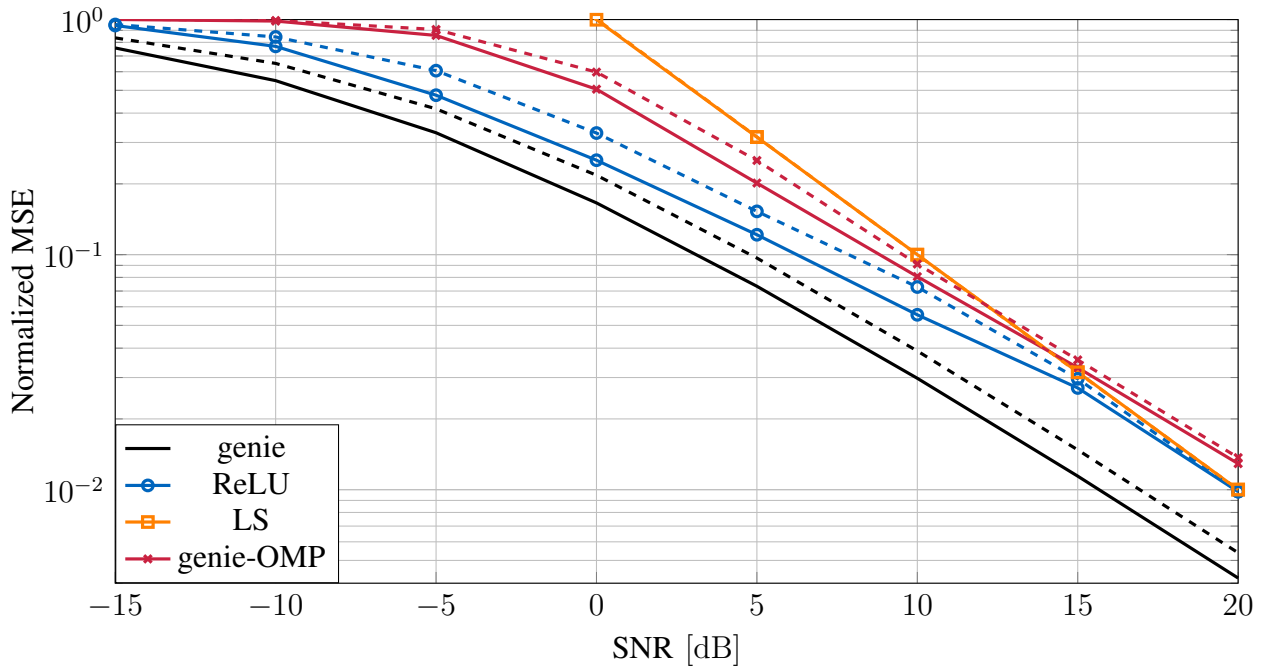
\begin{figure}[t]
	\centering
	\begin{tikzpicture}
%		\pgfplotsset{
%			compat=1.3,
%			tick label style={font=\scriptsize},
%			label style={font=\scriptsize},
%			legend style={font=\scriptsize}	
%		}
		\begin{axis}
			[width=1\columnwidth,
			height=0.55\columnwidth,
			xtick=data, 
			xmin=-15, 
			xmax=20,
			xlabel={SNR $[\operatorname{dB}]$},
			ymode = log, 
			ymin=4*1e-3,
			ymax=1,
			ylabel= {Normalized MSE}, 
			ylabel shift = 0.0cm,
			grid = both,
			legend columns = 1,
			legend entries={
				genie,
				ReLU,
				LS,
				genie-OMP,
			},
			legend style={at={(0.0,0.0)}, anchor=south west},
			]
			
			\addplot[mark options={solid},color=TUMBlack,line width=1.2pt]
			table[x= snr, y=nmse, col sep=comma]
			{csvdat/MIMO/downlink/n_paths_3_angle_spreadBS_2_angle_spreadMS_35_n_antennasBS_16_n_antennasMS_16_snr_min_-15.0_snr_max_20.0_unquant_16pilot.csv};
			
			\addplot[mark options={solid},color=TUMBlue,line width=1.2pt,mark=o]
			table[x= SNR, y=cnn_fft_relu_non_hier_False, col sep=comma]
			{csvdat/MIMO/downlink/2020-12-07_17-58-44_3paths_16BS_antennas_16MS_antennas_35AS_40lbatch_400ebatch_20lbatchsize_16pilots_16trainpilots.csv};
			
			\addplot[mark options={solid},color=TUMBeamerOrange,line width=1.2pt,mark=square]
			table[x= SNR, y=Genie_Aided, col sep=comma]		{csvdat/LS/2020-11-25_09-01-36_3paths_16BS_antennas_16MS_antennas_35AS_40lbatch_400ebatch_20lbatchsize_16pilots_4trainpilots.csv};
			
			\addplot[mark options={solid},color=TUMBeamerDarkRed,line width=1.2pt,mark=x]
			table[x= SNR, y=Genie_OMP, col sep=comma]		{csvdat/MIMO/downlink/2020-12-02_11-47-45_3paths_16BS_antennas_16MS_antennas_35AS_40lbatch_400ebatch_20lbatchsize_16pilots_16trainpilots.csv};
			
			\addplot[mark options={solid},color=TUMBlack,line width=1.2pt,dashed]
			table[x= snr, y=nmse, col sep=comma]
			{csvdat/MIMO/downlink/n_paths_3_angle_spreadBS_2_angle_spreadMS_35_n_antennasBS_8_n_antennasMS_8_snr_min_-15.0_snr_max_20.0_unquant_8pilot.csv};
			
			\addplot[mark options={solid},color=TUMBlue,line width=1.2pt,mark=o,dashed]
			table[x= SNR, y=cnn_fft_relu_non_hier_False, col sep=comma]		{csvdat/MIMO/downlink/2020-12-07_17-59-13_3paths_8BS_antennas_8MS_antennas_35AS_40lbatch_400ebatch_20lbatchsize_8pilots_8trainpilots.csv};
			
			\addplot[mark options={solid},color=TUMBeamerOrange,line width=1.2pt,mark=square,dashed]
			table[x= SNR, y=Genie_Aided, col sep=comma]		{csvdat/MIMO/downlink/2020-12-07_18-21-00_3paths_8BS_antennas_8MS_antennas_35AS_40lbatch_400ebatch_20lbatchsize_8pilots_8trainpilots.csv};
			\addplot[mark options={solid},color=TUMBeamerDarkRed,line width=1.2pt,mark=x,dashed]
			table[x= SNR, y=Genie_OMP, col sep=comma]		{csvdat/MIMO/downlink/2020-12-07_18-00-02_3paths_8BS_antennas_8MS_antennas_35AS_40lbatch_400ebatch_20lbatchsize_8pilots_8trainpilots.csv};
			
		\end{axis}
	\end{tikzpicture}
	\caption{Performance comparison of a $8\times 8$ (dashed) and $16\times 16$ (solid) MIMO setup with 3 clusters and $N=U$.}
	\label{fig:diff_mimo}
	\vspace{-0.5cm}
\end{figure}

In Fig. \ref{fig:diff_mimo} we also show different antenna setups, namely a $8\times 8$ and $16\times 16$ MIMO setting, where $U$=$N$. This illustrates the huge application area of the proposed approach, which is again able to outperform genie-\ac{omp} with a large performance gap, especially in the low to medium \ac{snr} region. The LS estimator achieves the same performance in both~settings.

\section{Conclusion}
We proposed a learning-based \ac{cnn} approach for estimating MIMO channels, which is a non-trivial generalization of the \ac{cnn} estimator from \cite{8272484}. We have shown that the high-level architecture with blockdiagonal structure simplifies to a diagonal structrue when using \ac{dft}-based pilot matrices and that the estimator breaks down to the SIMO estimator from \cite{8272484} for a single-antenna \ac{MS} and a single pilot. The online complexity of the \ac{ce} approach is only $\mathcal{O}(SU\log(SU))$ floating point operations which is equal to low-complexity approaches like \ac{omp} or the discussed \ac{ml} estimator. 
Simulation results depicted performance gains compared to state-of-the-art algorithms, especially for involved settings which model real-world~scenarios. 
%The performance of the proposed estimator was evaluated for different sceanrios, where it was able to outperform state-of-the-art algorithms, especially for involved settings which model real-world~scenarios. 

%We have proposed a learning-based \ac{cnn} approach for estimating MIMO channels, which is a non-trivial generalization of the \ac{cnn} estimator from \cite{8272484}, by incorporating model-based assumptions into the network structure and use learning to compensate for the mismatch of these assumptions in real scenarios. It was further shown that the general structure of the estimator simplifies drastically when using \ac{dft}-based pilot matrices which enable a low-complexity implementation. The performance of the proposed estimator was evaluated for different sceanrios and aspects, where it was able to compete and outperform state-of-the-art algorithms with comparable complexity, especially for very involved settings which model real-world scenarios. 

%\clearpage
\appendix
\subsection{Proof of the MMSE Filter Formulation}\label{app:lemma1}
The likelihood of $\bm{y}$ given $\bm{\delta}$ is assumed to be Gaussian:
\begin{align}
	p(\bm{y}|\bm{\delta}) &\propto \exp(-\bm{y}^\op{H}\bm{C}_{\bm{y}|\bm{\delta}}^{-1}\bm{y})|\bm{C}_{\bm{y}|\bm{\delta}}|\inv
	\\
	&\propto \exp(-\operatorname{tr}(\bm{C}_{\bm{y}|\bm{\delta}}^{-1}\bm{yy}^{\op{H}}))|\bm{C}_{\bm{y}|\bm{\delta}}^{-1}|.
\end{align}
Now we wish to express $\bm{C}_{\bm{y}|\bm{\delta}}^{-1}$ in terms of $\bm{W_\delta}$, which is similar to the problem in \cite{9145371} and can be computed as $\bm{C}_{\bm{y}|\bm{\delta}}^{-1} = \sigma^{-2}(\bm{\op{I}}_{SN} - \bm{XW_\delta})$.
The likelihood is now re-expressed as
\begin{align}
	p(\bm{y}|\bm{\delta}) 
	%	&\propto \exp(-\operatorname{tr}(\sigma^{-2}(\bm{\op{I}} - \bm{XW_\delta})\bm{yy}^{\op{H}}))|\sigma^{-2}(\bm{\op{I}}_{SN}-\bm{XW_\delta})| 
	%	\\
	&\propto \exp(\operatorname{tr}(\sigma^{-2} \bm{XW_\delta}\bm{yy}^{\op{H}}))|\bm{\op{I}}_{SN}-\bm{XW_\delta}|
	\\
	&\propto \exp(\operatorname{tr}(\bm{XW_\delta}\Chat) + b ),
	\label{eq:likelihood_pilots}
\end{align}
with $\Chat = \B y\B y\h$ and $b = \log|\eye_{SN} - \B X\Wdel|$.
%with the definition $b_{\bm{\delta}} = \log|\bm{\op{I}} - \bm{XW_\delta}|$.
%Since the noise covariance matrix does not depend on $\bm{\delta}$, we can further simplify the expression to
Plugging \eqref{eq:likelihood_pilots} into \eqref{eq:Wstar} gives the \ac{mmse} filter as in \eqref{eq:mmse} for a sample of $\bdelt$.

\subsection{The Diagblock Operator}\label{app:diagblock}
We define the operator $\diagblock_{S,U}(\B x)$ which takes a vector $\B x\in\mathbb{C}^{SU^2}$ as input and outputs a sparse block matrix $\B X\in\mathbb{C}^{SU\times SU} $ with $U\times U$ diagonal blocks of size $S\times S$. The diagonal entries of the blocks (in row-major block order) are then exactly the entries of the vector. For example:
\begin{equation}\label{eq:example}\nonumber
	\diagblock_{2,2}(\begin{bmatrix}x_1&x_2&\cdots &x_8\end{bmatrix}\T) = \begin{bmatrix}x_1 &0 &x_3&0\\ 0&x_2&0&x_4\\ x_5 &0&x_7&0\\0&x_6&0&x_8\end{bmatrix}.
\end{equation}
Equivalently, we define the case where the operator takes a block matrix $\B X\in\mathbb{C}^{SU\times SU}$ as input and outputs the vector $\B x\in\mathbb{C}^{SU^2}$, consisting of the block-diagonals of the matrix. 
%In other words, it performs the opposite direction from \eqref{eq:example}.

\subsection{Verification of Assumption 2}\label{app:SE}
%As described in Section \ref{sec:model}, the transmit and receive covariance matrices are Toeplitz. 
Based on the result from \cite[Appendix B]{CIT-006} and also discussed in \cite{8272484}, the  transmit and receive covariance matrices are Toeplitz and thus asymptotically equivalent to corresponding circulant matrices. Since all circulant matrices have the columns of the \ac{dft} matrix $\B F$ as eigenvectors, it holds that
\begin{equation}
	\B C_{\bdelt,\{T,R\}} \asymp \B F_{\{U,S\}}\h\diag(\B c_{\bdelt,\{T,R\}}) \B F_{\{U,S\}}.
\end{equation}
%where $\B c_{\bdelt,\{T,R\}}$ contains the diagonal elements of $\B F_{\{T,R\}} \B C_{\bdelt,\{T,R\}}\B F_{\{T,R\}}\h$.
Furthermore, the overall covariance matrix $\Cdel$ is asymptotically equivalent to a block-circulant matrix, shown by
\begin{align}
	\Cdel
	&\asymp \left(  \B F_U\h \diag(\B c_{\bdelt,T}) \B F_U\right)\otimes \left( \B F_S\h \diag(\B c_{\bdelt,R}) \B F_S \right)\\
	%	&=\B Q\h \left(\diag(\B c_{\bdelt,T})\otimes\diag(\B c_{\bdelt,R})\right) \B Q \\
	&=\B Q\h \diag(\B c_{\bdelt})\B Q.\label{eq:asymp_circ}
\end{align}
This is a good approximation for large-scale systems \cite{8272484}, \cite{https://doi.org/10.1002/cpa.20064}.
If we plug this asymptotic equivalency from \eqref{eq:asymp_circ} into the conditional \ac{mmse} from \eqref{eq:cond_mmse} by using the substitutions $\B D_{\bdelt} = \diag(\B c_{\bdelt})$ and $\tilde{\B Q} = \B Q\B X\h$, we get
\begin{equation}
	\Wdel = \B Q\h \B D_{\bdelt}\tilde{\B Q}(\tilde{\B Q}\h\B D_{\bdelt}\tilde{\B Q}+\sigma^2\eye)\inv.
\end{equation}
We can assume that $\diag(\B c_{\bdelt})$ is positive definite as $\B c_{\bdelt}$ contains the eigenvalues of $\Cdel$. Therefore, we can use the matrix inversion identity from \cite[Lemma 2]{Lecture} such that we get
\begin{equation}\label{eq:se_derivation}
	\Wdel = \B Q\h \sigma^{-2}(\sigma^{-2}\tilde{\B Q}\tilde{\B Q}\h+\B D_{\bdelt}\inv)\inv\tilde{\B Q}.
\end{equation}
Therein, the term $\tilde{\B Q}\tilde{\B Q}\h = \B Q\B X\h\B X\B Q$ is a block matrix with diagonal blocks, such that we get the expression as in \eqref{eq:ass2_se}.

If we then plug the expression from Assumption \ref{ass:2} into the \ac{ge} from \eqref{eq:ge}, we can reformulate the trace expression 
\begin{equation*}
	\begin{aligned}
		\tr(\siginv\tilde{ \B Q}\h\diagblock_{S,U}(\B w\deli)\tilde{\B Q}\B y\B y\h) 
		= 	\tr(\diagblock_{S,U}(\B w\deli)\B \Lambda)
	\end{aligned}\label{eq:trace}
\end{equation*}
by interpreting the matrix $\B \Lambda= \siginv\tilde{\B Q}\B y\B y\h\tilde{ \B Q}\h$ as a block matrix, consisting of $U\times U$ blocks of size $ S\times S$. The product inside the trace is then a sum of block products with the property of diagonal blocks in $\diagblock_{S,U}(\B w\deli)$. By exploiting the fact that the trace of the product of a diagonal and a arbitrary matrix is equal to the inner product of their diagonals, only the diagonals of the blocks in $\B \Lambda$ are needed for the calculation of the trace. This refers to the diagblock structure we defined in Appendix \ref{app:diagblock}, and the trace expression simplifies to 
\begin{equation}
	\tr(\diagblock_{S,U}(\B w\deli)\B \Lambda) = \B w\deli\T \diagblock_{S,U}(\B \Lambda).
\end{equation}
If we then rewrite the sums in \eqref{eq:ge} as matrix-vector products, we obtain the expression from \eqref{eq:SE_element}.

%\subsection{2D Shift Invariance}\label{app:ass3}
%In order to reasonably assume block-circulant matrices $\B A\SE$ in \eqref{eq:SE_element}, several assumptions are needed. In \cite[Appendix D]{8272484}, the shift invariance is discussed for the SIMO case with a \ac{ula} at the \ac{BS}. In the MIMO case for the spatial channel model, the covariance matrix can be decomposed as the Kronecker prodcut of the transmit and receive side covariance matrices. Therefore, the 1D shift invariance from \cite{8272484} for the single propagation path case in the 3GPP model extends to a 2D shift invariance and the matrix $\B A\SE$ can be approximated by 
%\begin{equation}
%	\B A\SE = \B Q \diag(\B Q \B w_0)\B Q.
%\end{equation}
%This is equal to constructing $\B A\SE$ by a 2D circular convolution with the uniform samples $\B w_0$ as convolution kernel.

%\subsection{Truncated DFT Pilot Matrix}\label{app:diag_se}
%If we have truncated \ac{dft} matrices which fulfill $\B X\h\B X = \frac{N}{U}\eye_{SU}$, then the matrix product $\tilde{\B Q}\tilde{\B Q}\h$ in \eqref{eq:se_derivation} becomes diagonal and Assumption 2 can be written in terms of \eqref{eq:se_diag}.
%\vspace{1cm}
%\begin{strip}
%	\hrule
%	\bigskip
%	this is a test
%\end{strip}
%~
%\clearpage 
%\twocolumn
\bibliographystyle{IEEEtran}
\bibliography{IEEEabrv,bibliography}

\end{document}